\documentclass[]{aastex631}
\usepackage{tabularx}
\usepackage{amsmath}
\usepackage{longtable}
\usepackage{multirow}

\usepackage{color,soul}
\soulregister\citet7
\soulregister\citep7
\soulregister\ref7
\soulregister\eqref7
\soulregister\url7

\renewcommand\hl[1]{#1}

\newcommand{\appropto}{\mathrel{\vcenter{
  \offinterlineskip\halign{\hfil$##$\cr
    \propto\cr\noalign{\kern2pt}\sim\cr\noalign{\kern-2pt}}}}}

\begin{document}

\title{JWST observations of K2-18b can be explained by a gas-rich mini-Neptune with no habitable surface}

\author[0000-0002-0413-3308]{Nicholas F. Wogan}
\affiliation{Space Science Division, NASA Ames Research Center, Moffett Field, CA 94035}
\affiliation{Virtual Planetary Laboratory, University of Washington, Seattle, WA 98195}

\author[0000-0003-1240-6844]{Natasha E. Batalha}
\affiliation{Space Science Division, NASA Ames Research Center, Moffett Field, CA 94035}

\author{Kevin Zahnle}
\affiliation{Space Science Division, NASA Ames Research Center, Moffett Field, CA 94035}
\affiliation{Virtual Planetary Laboratory, University of Washington, Seattle, WA 98195}

\author{Joshua Krissansen-Totton}
\affiliation{Department of Earth and Space Sciences, University of Washington, Seattle, WA 98195}
\affiliation{Virtual Planetary Laboratory, University of Washington, Seattle, WA 98195}

\author{Shang-Min Tsai}
\affiliation{University of California Riverside, Riverside, CA 92521}

\author[0000-0003-2215-8485]{Renyu Hu}
\affiliation{Jet Propulsion Laboratory, California Institute of Technology, Pasadena, CA 91109}
\affiliation{Division of Geological and Planetary Sciences, California Institute of Technology, Pasadena, CA 91125}

\begin{abstract}
  JWST recently measured the transmission spectrum of K2-18b, a habitable-zone sub-Neptune exoplanet, detecting CH$_4$ and CO$_2$ in its atmosphere. The discovery paper argued the data are best explained by a habitable ``Hycean'' world, consisting of a relatively thin H$_2$-dominated atmosphere overlying a liquid water ocean. Here, we use photochemical and climate models to simulate K2-18b as both a Hycean planet and a gas-rich mini-Neptune with no defined surface. We find that a lifeless Hycean world is hard to reconcile with the JWST observations because photochemistry only supports $< 1$ part-per-million CH$_4$ in such an atmosphere while the data suggest about $\sim 1\%$ of the gas is present. Sustaining \%-level CH$_4$ on a Hycean K2-18b may require the presence of a methane-producing biosphere, similar to microbial life on Earth $\sim 3$ billion years ago. On the other hand, we predict that a gas-rich mini-Neptune with $100 \times$ solar metallicity should have 4\% CH$_4$ and nearly 0.1\% CO$_2$, which are compatible with the JWST data. The CH$_4$ and CO$_2$ are produced thermochemically in the deep atmosphere and mixed upward to the low pressures sensitive to transmission spectroscopy. The model predicts H$_2$O, NH$_3$ and CO abundances broadly consistent with the non-detections. Given the additional obstacles to maintaining a stable temperate climate on Hycean worlds due to H$_2$ escape and potential supercriticality at depth, we favor the mini-Neptune interpretation because of its relative simplicity and because it does not need a biosphere or other unknown source of methane to explain the data.
\end{abstract}

\section{Introduction}

Whether or not life is abundant in the galaxy depends on the frequency of habitable worlds. The Kepler era of exoplanet exploration revealed that close-in sub-Neptunes ($\sim 2.4$ $R_\earth$) have high occurrence rates \citep{Fulton_2018}. These planets have bulk densities that can be explained by several planetary models that range from a massive H$_2$ atmosphere similar to Neptune's, to a thin hydrogen atmosphere (e.g., $\sim 1$ bar) overlying a H$_2$O-rich interior. Researchers have suggested that H$_2$O-rich sub-Neptunes could have habitable surface oceans provided that the climate is suitable for liquid water \citep{Madhusudhan_2021}. These so-called ``Hycean'' worlds, if they exist, have the potential to be among the most common habitable planetary environments.


Perhaps the best known Hycean world candidate is the sub-Neptune K2-18b \citep[8.63 M$_{\earth}$, 2.61 R$_{\earth}$;][]{Benneke_2019}), which was recently observed by the James Webb Space Telescope \citep[i.e., JWST;][]{Madhusudhan_2023}. The transmission spectrum reveals strong evidence for CH$_4$ and CO$_2$ in a H$_2$-rich atmosphere. Furthermore, JWST did not detect NH$_3$, H$_2$O or CO. \citet{Madhusudhan_2023} argued the data are best explained by a habitable Hycean world because, according to past photochemical studies, such a planet can be consistent with the NH$_3$ non-detection \citep{Tsai_2021,Yu_2021,Madhusudhan_2023_chem,Hu_2021}. Ammonia is instead expected on a mini-Neptune with a massive hydrogen atmosphere \citep[e.g.,][]{Yu_2021,Hu_2021}. Furthermore, \citet{Madhusudhan_2023} favored a Hycean world because their retrieved $\sim 1\%$ abundances for CH$_4$ and CO$_2$ are broadly compatible with photochemical modeling predictions made by \citet{Hu_2021} for a Hycean K2-18b.

Here, we use 1-D photochemical and climate models to revisit the past calculations \citep{Hu_2021,Yu_2021,Tsai_2021} that support a habitable ocean-world interpretation of the data. We simulate K2-18b as a Hycean planet to determine whether the CH$_4$ and CO$_2$ suggested by JWST are photochemically stable in such an atmosphere. Our Hycean models consider both a lifeless and inhabited planet, the latter represented by a primitive microbial biosphere that influence atmosphere chemistry. We also model K2-18b as a gas-rich mini-Neptune with a deep atmosphere. By comparing our simulations to the JWST data, and considering the relative complexities for each simulated composition, we suggest the most likely planetary model for K2-18b.

\section{Methods}

\subsection{Hycean worlds} \label{sec:methods_hycean}

To simulate a Hycean K2-18b, we first modeled a pressure-temperature (P-T) profile using the climate code contained within the \emph{Photochem} software package \citep{Wogan_2023}. The climate model uses correlated-k radiative transfer with opacities detailed in Appendix D of \citet{Wogan_2023}. The code constructs P-T profiles assuming the lower atmosphere follows a moist-pseudo adiabat connected to an isothermal stratosphere. For K2-18b, we assume a 215 K stratosphere following \citet{Hu_2021_gas}. Our approach can consider any number of condensing species \citep[e.g.,][]{Graham_2021}, but H$_2$O is the most important condensable for a habitable K2-18b.

\begin{figure*}
  \centering
  \includegraphics[width=0.85\textwidth]{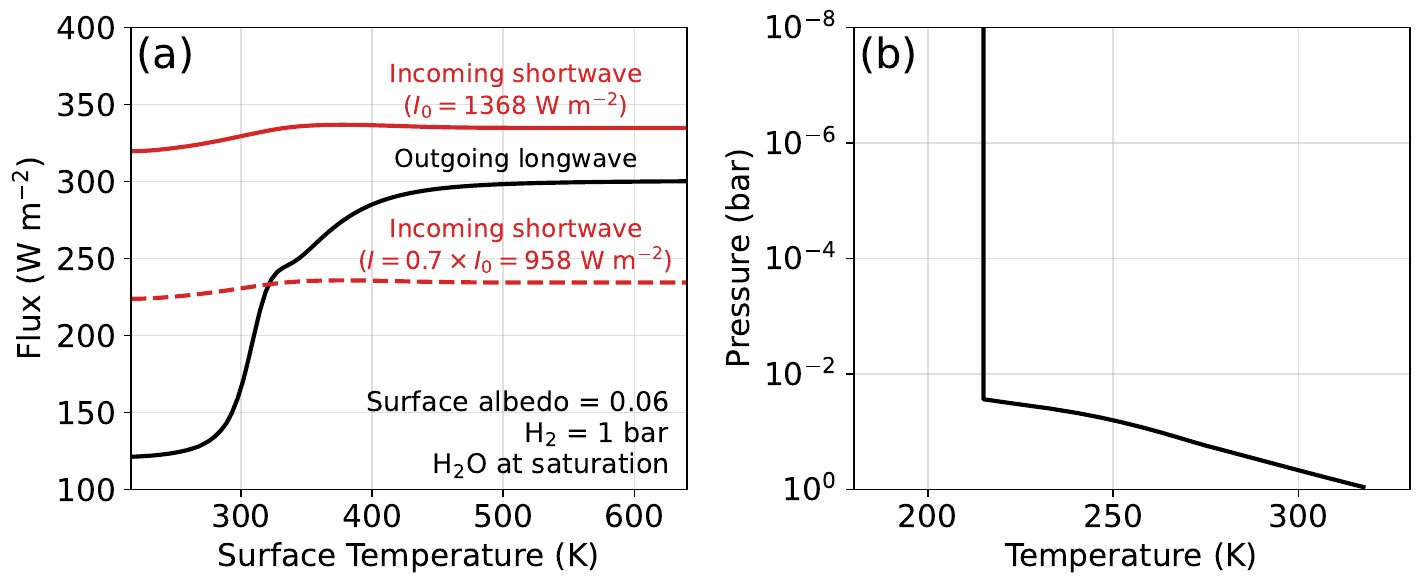}
  \caption{The climate of a plausible Hycean K2-18b. (a) shows the incoming shortwave (red) and outgoing longwave radiation (black) as a function of surface temperature computed with the climate model in \emph{Photochem}. Calculations assume 1 bar of H$_2$ with H$_2$O at saturation in the troposphere. The solid red line is the incoming shortwave radiation for K2-18b's full stellar insolation (1368 W m$^{-2}$), which can not be balanced by outgoing longwave energy below the critical point of H$_2$O ($< 647$ K). The dashed red line considers a 30\% smaller insolation to crudely represent high-altitude clouds reflecting away starlight, which would allow for a stable $\sim 320$ K climate. (b) the P-T profile for the stable climate in (a) which we adopt for the Hycean photochemical simulations.}
  \label{fig:olr_isr}
\end{figure*}

For a Hycean K2-18b, we nominally assume a 1-bar H$_2$-dominated atmosphere with a water-saturated troposphere to facilitate comparison with previous work \citep{Hu_2021,Madhusudhan_2023_chem,Innes_2023}. For such a composition, our cloud-free climate model predicts that K2-18b would not be habitable because the surface temperature would exceed the critical point of H$_2$O (Figure \ref{fig:olr_isr}a), consistent with past studies \citep{Innes_2023,Scheucher_2020}. However, it has been suggested that high-altitude clouds or hazes could potentially reflect short wave radiation allowing for a cooler climate \citep{Madhusudhan_2021,Piette_2020}. To approximate the cooling effects of clouds in our cloud-free climate simulations, following \citet{Hu_2021}, we arbitrarily reduce the incoming solar radiation by 30\% which permits a $\sim 320$ K surface (Figure \ref{fig:olr_isr}a). We adopt this habitable P-T profile, shown in Figure \ref{fig:olr_isr}b, for all Hycean scenarios. Our photochemical simulations include up to \%-level CH$_4$ and CO$_2$, but the Figure \ref{fig:olr_isr}b P-T profile ignores their greenhouse contribution. For this analysis, this is justified because the climate of a Hycean K2-18b is uncertain, and our climate model predicts the surface temperature would increase by $\lesssim 10$ K when accounting for CH$_4$ and CO$_2$.

With our estimated P-T profile (Figure \ref{fig:olr_isr}b), we then simulate steady-state photochemistry using \emph{Photochem} \citep{Wogan_2024_photochem}. The photochemical model contained in \emph{Photochem} solves a system of partial differential equations approximating molecular transport in the vertical direction and the effect of chemical reactions, photolysis and condensation. We have made several updates to the reaction network and thermodynamic data originally published in \citet{Wogan_2023}. We improved the kinetics and thermodynamics of $\mathrm{CH_3O}$, $\mathrm{H_2COH}$ and related species, all of which are detailed in Appendix Table \ref{tab:updated_reactions}. For key reactions, we nominally adopt new kinetics following \citet{Xu_2015}, but we also consider alternative rates from \citet{Klippenstein_priv_com}. Theses updates are important for estimating photochemical methane production on Hycean worlds as discussed in Section \ref{sec:result_hycean}. We have also updated our H$_2$O and H$_2$ photolysis data (Appendix Table \ref{tab:updated_reactions}). The updated network is available on Zenodo (see the ``.yaml'' files in the ``input/'' folder of \citet{Wogan_2024_reproduce}). Because the UV spectrum of K2-18 has not been measured, we instead use the UV spectrum of GJ 176 measured by the MUSCLES survey for our photochemical calculations \citep{France_2016} following the \citet{Hu_2021} analysis.

\begin{table*}
  \caption{Model scenarios}
  \label{tab:models}
  \begin{center}
  \begin{tabularx}{.96\linewidth}{p{0.15\linewidth} | p{0.07\linewidth} | p{0.07\linewidth} | p{0.085\linewidth} | p{0.099\linewidth} p{0.099\linewidth} p{0.099\linewidth} p{0.14\linewidth} }
    \hline \hline
    \multirow{2}{*}{Model type} & \multirow{2}{*}{Model \#} & \multirow{2}{*}{$K_{zz}$$^\text{a}$} & \multirow{2}{*}{Metallicity} & \multicolumn{4}{c}{Lower boundary condition$^\text{d}$} \\
    & & & & N$_2$ & CO$_2$ & CH$_4$ & CO \\
    \hline
    \multirow{1}{*}{Lifeless Hycean$^\text{b}$} & 1 & $5 \times 10^{5}$ & - & $f = 3 \times 10^{-3}$ & $f = 8 \times 10^{-3}$ & $\Phi = 0$ & $\Phi = 0$ \\
    Inhabited Hycean$^\text{b}$ & 2 & $5 \times 10^{5}$ & - & $f = 3 \times 10^{-3}$ & $f = 8 \times 10^{-3}$ & $\Phi = 5 \times 10^{10}$ & $v_\mathrm{d} = 1.2 \times 10^{-4}$ \\
    Mini-Neptune$^\text{c}$ & 3 & Figure \ref{fig:neptune_clima_comp}b & $100\times$ solar & \multicolumn{4}{c}{$f = $ chemical equilibrium$^\text{e}$} \\
    \hline
    \multicolumn{8}{p{0.94\linewidth}}{
      $^\text{a}$The vertically-constant eddy diffusion coefficient in cm$^2$ s$^{-1}$

      $^\text{b}$All Hycean models include $7 \times 10^{-3}$ cm s$^{-1}$ deposition velocities for HCN and HCCCN \citep{Wogan_2023}, and impose a $10^{-5}$ cm s$^{-1}$ deposition velocity for C$_2$H$_6$ \citep{Hu_2021}.

      $^\text{c}$The mini-Neptune case has a solar C/O ratio and a 60 K intrinsic temperature.

      $^\text{d}$The variable $f$ indicates a fixed lower-boundary mixing ratio, $\Phi$ indicates a fixed surface flux in molecules cm$^{-2}$ s$^{-1}$, and $v_\mathrm{d}$ indicates a surface deposition velocity in cm s$^{-1}$. If a fixed surface flux is specified, then the deposition velocity is zero. \hl{In Hycean simulations,} unspecified molecules have a zero-flux lower boundary condition.

      $^\text{e}$\hl{In the mini-Neptune case, we assume fixed lower boundary conditions at chemical equilibrium for molecules with equilibrium concentrations $> 10^{-8}$ mixing ratio. For lower concentration molecules, we permit molecules to mix into the deep atmosphere ($> 500$ bar) with a deposition velocity $v_\mathrm{d} = K_{zz}/H$, where $H$ is scale height, following past works \citep{Moses_2000}.}
    }
  \end{tabularx}
  \end{center}
\end{table*}

\subsection{Mini-Neptune world} \label{sec:methods_neptune}

We additionally model K2-18b as a gas-giant mini-Neptune with no habitable surface. We take the same approach as \citet{Hu_2021_gas} and simulate the massive hydrogen atmosphere over two stages: The first considers the deep atmosphere (500 to 1 bar), and the second simulates the upper atmosphere (1 to $10^{-8}$ bar). The lower atmosphere stage captures the equilibrium-to-disequilibrium transition (i.e., gas quenching) that occurs deep in a gas-giant atmosphere \citep[e.g.,][]{Zahnle_2016}, while the upper atmosphere model approximates the impact of UV photolysis and gas condensation on composition.

For the first stage, we use the \emph{PICASO} climate model \citep{Mukherjee_2023} to generate a P-T profile with opacities appropriate for a $100 \times$ solar metallicity with a solar C/O at chemical equilibrium assuming a geothermal heat flow consistent with an intrinsic temperature ($T_\mathrm{int}$) of 60 K \citep{Hu_2021_gas}. Note that the intrinsic temperature affects the upper atmosphere abundance of gases such as CH$_4$ \citep{Fortney_2020}. We discuss this $T_\mathrm{int}$ dependence in Section \ref{sec:result_neptune}, but leave a full parameter space exploration for future work. Next, using the P-T profile, we do a full kinetics simulation with the \emph{Photochem} model between 500 bar and 1 bar using our network of $\sim 600$ reversible reactions described previously (Section \ref{sec:methods_hycean}). We fix the lower boundary to chemical equilibrium composition, and allow the kinetics model to predict the chemical equilibrium-to-disequilibrium transition. The deep atmosphere adopts an altitude-independent eddy diffusion coefficient of $K_{zz} = 10^{8}$ cm$^{2}$ s$^{-1}$ following \citet{Hu_2021_gas}.

In the second stage we simulate K2-18b's upper atmosphere using results from the first stage as lower boundary conditions. We do not use the \emph{PICASO} P-T profile above 1 bar because \emph{PICASO} assumes the entire atmospheric profile is at chemical equilibrium which would not be the case for the cool upper atmosphere of K2-18b. The chemical equilibrium assumption creates a stratospheric inversion in the P-T profile from greenhouse gases such as CH$_4$. Furthermore, \emph{PICASO} assumes a dry convective lapse rate but the P-T profile in much of the upper troposphere should follow a moist-pseudo adiabat because of water condensation. As an alternative to \emph{PICASO}, we extrapolate the P-T profile above the 1 bar level by drawing an adiabat upwards using the \emph{Photochem} climate model until it intersects an isothermal 215 K stratosphere. Appendix Figure \ref{fig:neptune_clima_comp}a compares the \emph{PICASO} profile to the modified profile that we adopt. Finally, using the modified P-T profile, we compute the photochemical steady-state of the upper atmosphere (1 to $10^{-8}$ bar) to predict its composition. At the lower boundary, we fix all gas concentrations to the values predicted at the 1 bar level of the lower atmosphere kinetics simulation described in the previous paragraph. The upper atmosphere simulation assumes a Jupiter-like eddy diffusion profile as used in \citet{Hu_2021_gas} (Appendix Figure \ref{fig:neptune_clima_comp}b).

\subsection{Transmission spectra}

We use the \emph{PICASO} code \citep{Batalha_2019} to compute the transmission spectra of simulated Hycean and mini-Neptune atmospheres adopting the R=60,000 resampled opacities archived on Zenodo \citep{Batalha_2022_opacities}. The main text presents clear-sky spectra because the JWST data do not favor high altitude clouds. The Appendix shows the spectral effects of water, elemental sulfur (S$_2$ and S$_8$) and hydrocarbon clouds and hazes.

\section{Results}

\subsection{Hycean worlds} \label{sec:result_hycean}

To investigate K2-18b as a Hycean world we first consider an uninhabited planet. Our nominal simulation, called ``model 1'', assumes a 1 bar H$_2$-dominated atmosphere, 0.8\% CO$_2$ fixed at the surface, and other settings and boundary conditions detailed in Table \ref{tab:models}. We choose 0.8\% CO$_2$ because it is the median concentration implied by the JWST spectrum \citep{Madhusudhan_2023}. Methane has a zero-flux lower boundary condition, therefore all accumulated CH$_4$ is the result of the photochemical reduction of CO$_2$ to CH$_4$. Figure \ref{fig:habitable_mix_p}a shows the steady-state composition of the model 1 atmosphere as a function of pressure revealing that only 0.4 part-per-billion (ppb) CH$_4$ is photochemically stable. Methane is slowly produced by the following sequence of reactions:
\begin{figure*}
  \centering
  \includegraphics[width=0.85\textwidth]{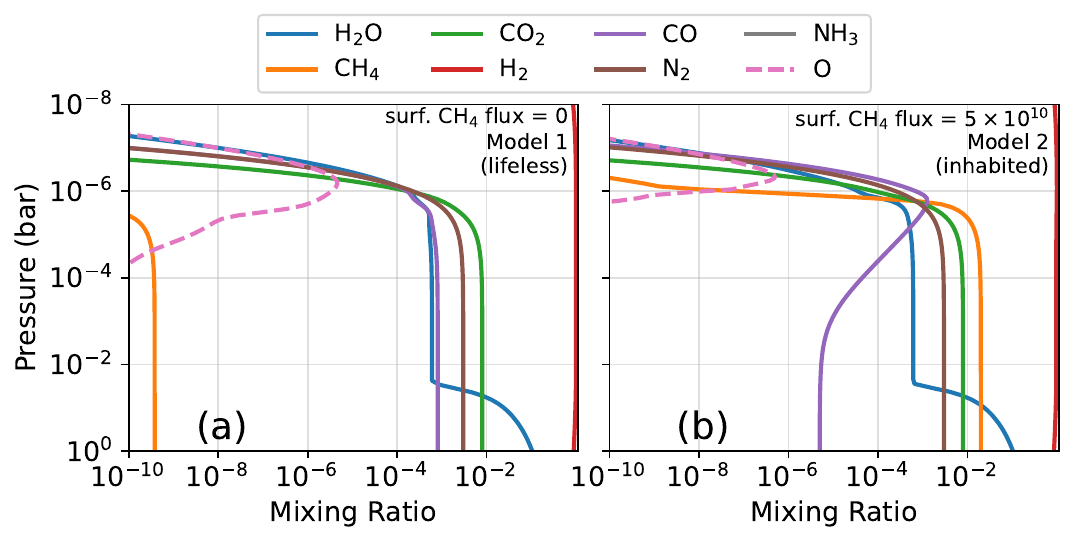}
  \caption{Photochemical simulations of K2-18b as a habitable Hycean world. Panels (a) and (b) correspond to Model 1 and 2, respectively, described in Table \ref{tab:models}. Both panels include the surface CH$_4$ flux in molecules cm$^{-2}$ s$^{-1}$ required to sustain the CH$_4$ concentration. (a) shows that only ppb-level methane can accumulate photochemically (i.e., abiotically). (b) shows that methane is predicted to build-up to \%-levels assuming a biological surface CH$_4$ flux of $5 \times 10^{10}$ molecules cm$^{-2}$ s$^{-1}$ which is about half of the modern Earth's biological flux \citep{Jackson_2020}.}
  \label{fig:habitable_mix_p}
\end{figure*}
\begin{subequations} \label{eq:ch4_prod}
\begin{align}
  \mathrm{CO_2} + h\nu &\rightarrow \mathrm{CO} + \mathrm{O(^1D)} \label{eq:ch4_prod_1} 
  \\
  2\:\mathrm{H} + 2\:\mathrm{CO} + 2\:\mathrm{M} &\rightarrow 2\:\mathrm{HCO} + 2\:\mathrm{M} \label{eq:ch4_prod_2}
  \\
  \mathrm{HCO} + \mathrm{HCO} &\rightarrow \mathrm{H_2CO} + \mathrm{CO} \label{eq:ch4_prod_3}
  \\
  \mathrm{H} + \mathrm{H_2CO} + \mathrm{M} &\rightarrow \mathrm{CH_3O} + \mathrm{M} \label{eq:ch4_prod_4} 
  \\
  \mathrm{CH_3O} + \mathrm{H} &\rightarrow \mathrm{CH_3} + \mathrm{OH} \label{eq:ch4_prod_5} 
  \\
  \mathrm{CH_3} + \mathrm{H} + \mathrm{M} &\rightarrow \mathrm{CH_4} + \mathrm{M} \label{eq:ch4_prod_6}
  \\
  2\:\mathrm{H_2} + 2\:\mathrm{OH} &\rightarrow 2\:\mathrm{H_2O} + 2\:\mathrm{H} \label{eq:ch4_prod_7}
  \\
  \mathrm{O(^1D)} + \mathrm{H_2} &\rightarrow \mathrm{OH} + \mathrm{H} \label{eq:ch4_prod_8}
  \\
  \cline{1-2}
  \mathrm{CO_2} + 2\:\mathrm{H} + 3\:\mathrm{H_2} &\rightarrow \mathrm{CH_4} + 2\:\mathrm{H_2O} \tag{\ref*{eq:ch4_prod}, net}
\end{align}
\end{subequations}
By analyzing column-integrated reaction rates we have determined that Reaction \eqref{eq:ch4_prod_4} is the rate-limiting step. Another important path with the same net reaction replaces both reactions involving CH$_3$O with alternatives that depend on the isomer H$_2$COH:
\begin{align}
  \mathrm{H} + \mathrm{H_2CO} + \mathrm{M} &\rightarrow \mathrm{H_2COH} + \mathrm{M} \label{eq:ch4_prod2_4} 
  \\
  \mathrm{H_2COH} + \mathrm{H} &\rightarrow \mathrm{CH_3} + \mathrm{OH} \label{eq:ch4_prod2_5} 
\end{align}

Methane is effectively destroyed by photolysis followed by several oxidizing reactions:
\begin{subequations} \label{eq:ch4_dest}
\begin{align}
  \mathrm{CH_4} + h\nu &\rightarrow \mathrm{CH_3} + \mathrm{H} \label{eq:ch4_dest_1} 
  \\
  \mathrm{H_2O} + h\nu &\rightarrow \mathrm{O} + \mathrm{H} + \mathrm{H} \label{eq:ch4_dest_1_1} 
  \\
  \mathrm{CH_3} + \mathrm{O} &\rightarrow \mathrm{H_2CO} + \mathrm{H} \label{eq:ch4_dest_2} 
  \\
  \mathrm{H_2CO} + h\nu &\rightarrow \mathrm{CO} + \mathrm{H_2} \label{eq:ch4_dest_3} 
  \\
  \mathrm{CO} + \mathrm{OH} &\rightarrow \mathrm{CO_2} + \mathrm{H} \label{eq:ch4_dest_4} 
  \\
  \mathrm{H_2O} + h\nu &\rightarrow \mathrm{OH} + \mathrm{H} \label{eq:ch4_dest_5} 
  \\
  \cline{1-2}
  \mathrm{CH_4} + 2\:\mathrm{H_2O} &\rightarrow 6\:\mathrm{H} + \mathrm{H_2} + \mathrm{CO_2} \tag{\ref*{eq:ch4_dest}, net}
\end{align}
\end{subequations}
In H$_2$-rich solar system atmospheres (e.g., Saturn's), methane has a long lifetime to destruction because, after photolysis (Reaction \eqref{eq:ch4_dest_1}), it recombines: $\mathrm{CH_3} + \mathrm{H} + \mathrm{M} \rightarrow \mathrm{CH_4} + \mathrm{M}$ \citep{Moses_2000}. The same recombination is inefficient in model 1 because, unlike the gas-giants in the solar system, model 1 has substantially more oxidizing gases like H$_2$O. In particular, water vapor photolysis at Ly-$\alpha$ wavelengths ($\lambda = 126.56$ nm) produces oxygen atoms \citep[Reaction \eqref{eq:ch4_dest_1_1},][]{Slanger_1982} that rapidly oxidize CH$_3$ before CH$_4$ is reformed. \hl{Methyl is also destroyed by atomic oxygen sourced from a sequence of reactions involving CO$_2$ photolysis: $\mathrm{CO_2} + h\nu \rightarrow \mathrm{CO} + \mathrm{O}$, $\mathrm{H_2O} + h\nu \rightarrow \mathrm{H} + \mathrm{OH}$, and $\mathrm{CO} + \mathrm{OH} \rightarrow \mathrm{CO_2} + \mathrm{H}$ which has the net reaction $\mathrm{H_2O} + h\nu \rightarrow \mathrm{O} + \mathrm{H} + \mathrm{H}$.} Overall, efficient methyl oxidation in additional to slow CH$_4$ production (Reaction path \eqref{eq:ch4_prod}), results in only trace amounts of atmospheric CH$_4$.

Our result that CH$_4$ cannot accumulate in model 1 is not sensitive to many model assumptions. For example, we have recomputed model 1 with vertically constant $K_{zz}$ between $10^4$ and $10^6$ cm$^2$ s$^{-1}$, N$_2$ concentrations ($f_\mathrm{N_2}$) between $\sim 1$ ppm and 1\%, and troposphere relative humidities ($\phi$) spanning 0.1 to 1. Within this parameter space, our photochemical code predicts the maximum stable CH$_4$ concentration is only 4 ppb for $K_{zz} = 10^4$ cm$^2$ s$^{-1}$, $f_\mathrm{N_2} = 1$ ppm, and $\phi = 1$. As an additional test, we recomputed model 1 using alternative rates for Reactions \ref{eq:ch4_prod_4} and \ref{eq:ch4_prod2_4} derived by \citet{Klippenstein_priv_com} using ab initio methods. \citet{Klippenstein_priv_com} predicts that these important rate-limiting reactions are faster than the \citet{Xu_2015} rates we nominally assume (Appendix Table \ref{tab:updated_reactions}) at the temperatures and pressures relevant to Hycean atmospheres. Despite this difference, our model using the \citet{Klippenstein_priv_com} rates predicts only 32 ppb CH$_4$.

Up to this point, we have modeled K2-18b as a habitable, yet uninhabited planet. Now we consider an inhabited case, which we refer to as model 2. Model 1 imposes the surface concentration of H$_2$, CO$_2$ and N$_2$, but most all other gases, including CH$_4$ and CO, are dictated by photochemistry. If K2-18b is a Hycean world inhabited by microbial life then CH$_4$ and CO could be biologically modulated gases like they were on the anoxic Archean Earth \citep{Kharecha_2005,Wogan_2020,Thompson_2022}. Chemosynthetic methanogens can consume H$_2$ and CO$_2$ for energy, producing methane as a waste gas:
\begin{equation}
  \mathrm{CO_2} + 4\:\mathrm{H_2} \rightarrow \mathrm{CH_4} + 2\:\mathrm{H_2O}
\end{equation}
CO is also food for acetogenic microbes:
\begin{equation} \label{eq:co_consume}
  4\:\mathrm{CO} + 2\:\mathrm{H_2O} \rightarrow 2\:\mathrm{CO_2} + \mathrm{CH_3COOH}
\end{equation}
The produced $\mathrm{CH_3COOH}$ could have been food for acetotrophic methanotrophs ($\mathrm{CH_3COOH} \rightarrow \mathrm{CH_4} + \mathrm{CO_2}$). Model 2 simulates K2-18b as a Hycean world with boundary conditions representing biological influence from these early Archean metabolisms (Table \ref{tab:models}). To model methanogenic life, we impose a surface CH$_4$ flux needed to replicate the \%-level concentration implied by the JWST data, which ended up being half the modern Earth's biological methane flux \citep[$5 \times 10^{10}$ molecules cm$^{-2}$ s$^{-1}$,][]{Jackson_2020}. We also add a CO deposition velocity of $1.2 \times 10^{-4}$ cm s$^{-1}$ to approximate the influence of CO-consuming acetogens \citep[Reaction \eqref{eq:co_consume},][]{Kharecha_2005}. At photochemical steady-state, model 2 has 2\% CH$_4$ and a $\sim 10^{-5}$ CO mixing ratio at the surface (Figure \ref{fig:habitable_mix_p}b). 

With a methanogenic biosphere, CH$_4$ can accumulate to the \%-levels suggested by recent JWST observations \citep{Madhusudhan_2023}. In contrast, on an uninhabited Hycean K2-18b, CH$_4$ should be at only ppb-levels (model 1) because large concentrations can not accumulate photochemically and other non-biological sources of methane seem implausible (Section \ref{sec:discussion}).

\subsection{Mini-Neptune world} \label{sec:result_neptune}

\begin{figure*}
  \centering
  \includegraphics[width=0.8\textwidth]{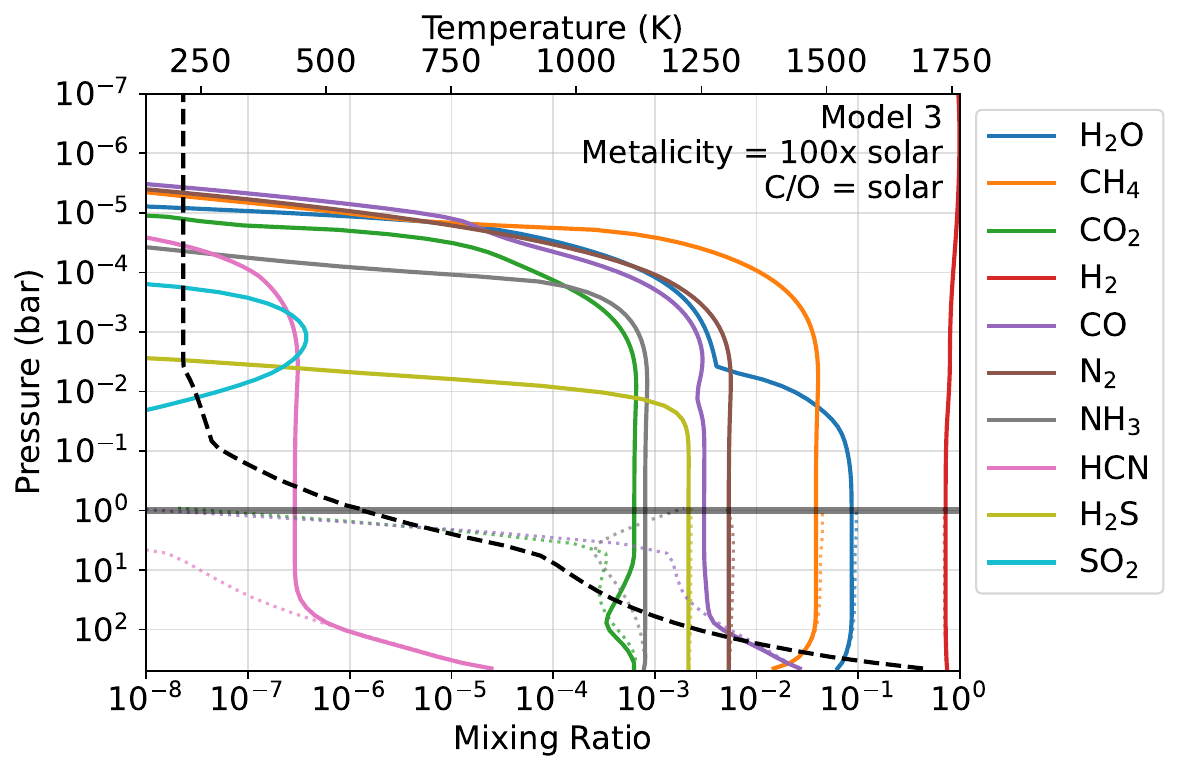}
  \caption{Climate and photochemical simulation of K2-18b as a mini-Neptune with no habitable surface (Model 3 in Table \ref{tab:models}). The black dashed line is the computed P-T profile which is referenced to the upper x-axis. The horizontal grey line at 1 bar divides the lower and upper atmosphere as discussed in Section \ref{sec:methods_hycean}. Solid colored lines are predicted atmospheric composition from our photochemical model. For comparison, the dotted lines in the lower atmosphere are chemical equilibrium composition. If K2-18b is a $100 \times$ solar mini-Neptune with a solar C/O ratio, then we predict the observable upper atmosphere should have $\sim 4\%$ CH$_4$ and nearly 0.1\% CO$_2$, which is in reasonable agreement with recent JWST observations \citep{Madhusudhan_2023}.}
  \label{fig:neptune_mix_t_p}
\end{figure*}

Figure \ref{fig:neptune_mix_t_p} shows K2-18b modeled as a gas-giant mini-Neptune with no habitable surface (i.e., model 3 in Table \ref{tab:models}). Deep in the atmosphere, at 500 bar and 1700 K, fast reactions enforce chemical equilibrium for our assumed composition of $100 \times$ solar metallicity with a solar C/O ratio. As gases mix upward to lower pressures and temperatures, reactions slow, causing an equilibrium-to-disequilibrium transition (i.e., gas quenching). N$_2$ and NH$_3$ chemistry quenches near 200 bar and $\sim 1400$ K, and the CO$_2$-CO-CH$_4$ system fails to maintain equilibrium near 100 bar and $\sim 1250$ K. These quench points are broadly consistent with \citet{Hu_2021_gas} who constructed similar mini-Neptune models of K2-18b. 

Quenched gases from the deep atmosphere mix upward to Model 3's stratosphere where they are relevant to transmission spectroscopy. The atmosphere has 4\% CH$_4$ along with 0.06\% CO$_2$ at 1 mbar, which is broadly consistent with recent JWST observations \citep{Madhusudhan_2023}. Water vapor condensation between 0.07 bar and 4 mbar reduces its concentration, causing only 0.3\% of the gas to be present at 1 mbar. At the same pressure level, there is also 0.3\% CO and 0.07\% NH$_3$. Only trace photochemically produced SO$_2$ is present ($\sim  10^{-7}$ mixing ratio) as most all sulfur is photochemically processed to S$_2$ and S$_8$ in the lower atmosphere where it condenses out \citep{Zahnle_2016}.

We additionally tested the sensitivity of Model 3 to the assumed intrinsic temperature ($T_\mathrm{int} = 60$ K), as this parameter can impact deep atmosphere quenching and the resulting stratospheric abundances of CH$_4$, CO$_2$, and CO \citep{Fortney_2020,Tsai_2021_comparative}. Larger $T_\mathrm{int}$ (e.g, 100 K) drives an increase in CO that is hard to reconcile with JWST observations. For lower $T_\mathrm{int}$ values (e.g., 30 K), our model does not produce enough CO$_2$ to explain the JWST data. Future abundance constraints from JWST offer an exciting avenue to study K2-18b's internal temperature and thermal evolution. We leave detailed exploration of this topic to a future study.

\subsection{Transmission spectra and comparison to JWST data}

\begin{figure*}
  \centering
  \includegraphics[width=1\textwidth]{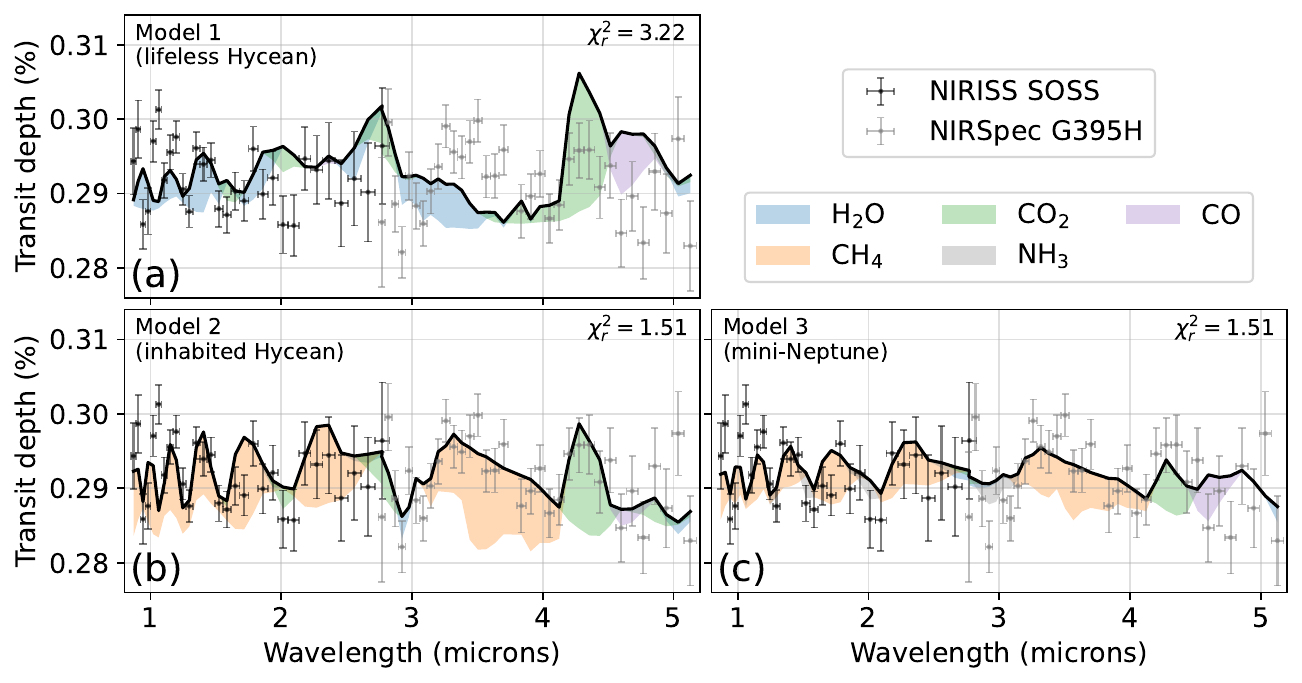}
  \caption{Transmission spectra of Hycean and mini-Neptune models of K2-18b compared to JWST NIRISS and NIRSpec data from Figure 3 in \citet{Madhusudhan_2023}. Panels (a), (b) and (c) show simulated clear-sky transmission spectra of models 1, 2 and 3, respectively. Colored shading shows the effect of molecules on the spectrum. The reported $\chi^2_r$ values each have 64 degrees of freedom. The JWST data strongly rules out a lifeless Hycean world (model 1, $\chi^2_r = 3.22$), but allows for either an inhabited Hycean (model 2) or mini-Neptune (model 3) model of K2-18b ($\chi^2_r = 1.51$).}
  \label{fig:spectra_clear_1}
\end{figure*}

Figure \ref{fig:spectra_clear_1} shows the simulated clear-sky transmission spectra of three scenarios for K2-18b compared to JWST NIRISS and NIRSpec observations: A lifeless Hycean planet (model 1), a Hycean world inhabited by an Archean-like biosphere (model 2) and a $100 \times$ solar metallicity mini-Neptune with no habitable surface (Model 3). In all cases, we allow the simulated spectra to have an offset between the NIRISS and NIRSpec data as to best fit the observations, motivated by \citet{Madhusudhan_2023}.

JWST data rules out model 1 ($\chi_r^2 = 3.22$) because the lifeless Hycean world does not have enough methane ($\sim 0.8$ ppb, Figure \ref{fig:habitable_mix_p}) to explain the observed CH$_4$ absorption shortwards of $4$ $\mu$m. On the other hand, the data do not strongly exclude an inhabited Hycean world (model 2, $\chi_r^2 = 1.51$). Model 2 fits the CH$_4$ and CO$_2$ spectral features in the data because it has 2\% of biologically-produced methane along with 0.8\% CO$_2$.

However, an inhabited Hycean world is not required to explain the data. Our model of a gas-giant mini-Neptune (model 3) has a comparable fit ($\chi_r^2 = 1.51$) largely because of its 4\% CH$_4$ and 0.06\% CO$_2$ at $\sim 1$ mbar. The spectra show small H$_2$O absorption because water vapor is cold trapped at $\sim 4$ mbar (Figure \ref{fig:neptune_mix_t_p}). Also, NH$_3$ has small absorption features at 1.5, 2, and 3 $\mu$m. \citet{Madhusudhan_2023} used the JWST data to argue for an NH$_3$ upper bound of $\sim 3 \times 10^{-5}$ at 95\% confidence assuming a vertically constant NH$_3$ concentration. At 1 mbar, our mini-Neptune model has $7 \times 10^{-4}$ NH$_3$, but photolysis rapidly diminishes the gas's concentration towards lower pressures (see Figure \ref{fig:neptune_mix_t_p}). Using a transmission contribution function (Equation 8 in \citet{Molliere_2019}) we find that the 3 $\mu$m NH$_3$ feature (Figure \ref{fig:spectra_clear_1}c) is sensitive to pressures between $\sim 10^{-3}$ and $\sim 10^{-5}$ bar where the ammonia concentration is between $7 \times 10^{-4}$ and $10^{-14}$ mixing ratio. While a direct comparison is challenging, our modeled heterogeneous NH$_3$ profile appears broadly compatible with the vertically constant upper bound derived by \citet{Madhusudhan_2023}.



Models 2 and 3, and the retrievals presented in \citet{Madhusudhan_2023}, are not able to reproduce the apparent large absorption feature near $\sim 1$ $\mu$m (red points in Appendix Figure \ref{fig:spectra_cloud}). Disregarding these six data points reduces $\chi_r^2$ values for both model 2 and 3 to about $1$. Future visits with NIRISS SOSS will be valuable to determine whether the scatter near $\sim 1$ $\mu$m is physical or instrumental.  

Our conclusion that the data strongly rule out model 1 but not model 2 or 3 is unchanged when including various aerosol opacities (Appendix Figure \ref{fig:spectra_cloud}). Accounting for aerosols, the model 1 simulated spectrum remains a poor fit ($\chi_r^2 = 2.33$) when compared to model 2 and 3 ($\chi_r^2 \approx 1.4$).

\section{Discussion} \label{sec:discussion}

\subsection{Reconciling our lifeless Hycean model with past research}

\citet{Hu_2021} pioneered the use of photochemical models to simulate K2-18b as a lifeless Hycean planet. They consider a 1 bar H$_2$ atmosphere, 1\% N$_2$, a comparable P-T profile to ours (Figure \ref{fig:olr_isr}b), and CO$_2$ concentrations between 400 ppm and 10\%. In all scenarios, they predict the photochemical accumulation of \%-levels CH$_4$, which contrasts with the ppb-level CH$_4$ we compute in very similar scenarios (e.g., model 1). There are two reasons our results differ. First, \citet{Hu_2021} assumed that the photolysis of H$_2$O produces only OH + H. However, we demonstrate here that the seemingly minor channel that produces O + H + H (Reaction \ref{eq:ch4_dest_1_1}) is important for CH$_4$ destruction in K2-18b's atmosphere. To verify this insight, we have rerun the 400 ppm-CO$_2$ model of \citet{Hu_2021} using their photochemical network and code, and with the sole inclusion of the O + H + H channel the steady-state CH$_4$ drops from 1\% to $3\times10^{-5}$ mixing ratio.

The second reason our results differ has to do with CH$_4$ production. \citet{Hu_2021} modeled Reaction \eqref{eq:ch4_prod_4}, a critical step to methane formation, with its high-pressure limit rate constant. This approach can accurately estimate the rate at high pressures ($\sim 100$ bar) but substantially over predicts the rate at $<1$ bar where third body collisions are more scarce. We have also rerun the 400 ppm-CO$_2$ model of \citet{Hu_2021} using their code, now also with the \citet{Xu_2015} pressure-dependent rate constant in Appendix Table \ref{tab:updated_reactions}, and find that the steady-state CH$_4$ mixing ratio further drops to $2\times10^{-7}$. This concentration is broadly consistent with model 1 shown in Section \ref{sec:result_hycean}, given remaining subtle differences in the temperature, diffusivity, and radiative transfer.


To further test the above explanation, we have also used the \emph{Photochem} code to reproduce the 400 ppm-CO$_2$ case in \citet{Hu_2021}. Adopting their boundary conditions, P-T profile, and eddy diffusion profile, our chemical network predicts $3 \times 10^{-7}$ mixing ratio CH$_4$ at steady state. When we perform the same simulation but use the \citet{Hu_2021} pressure-independent rate for Reaction \eqref{eq:ch4_prod_4}, our code predicts \hl{$10^{-4}$ mixing ratio} CH$_4$ should accumulate. \hl{When \emph{Photochem} also omits the O + H + H branch of H$_2$O photolysis, the CH$_4$ concentration further rises to $\sim 0.5\%$.} These \emph{Photochem} results are generally compatible with our \citet{Hu_2021} code calculations for the same experiments.

Furthermore, \citet{Madhusudhan_2023_chem} was unable to reproduce \citet{Hu_2021} using an independent photochemical model and network. ``Case 11'' in \citet{Madhusudhan_2023_chem} is very similar to the 10\%-CO$_2$ case in \citet{Hu_2021}, representing a 1 bar uninhabited Hycean world (i.e., zero-flux boundary conditions for CH$_4$ and CO). At photochemical steady-state, \citet{Madhusudhan_2023_chem} finds that only 55 ppb CH$_4$ should persist, which aligns with our conclusion that methane should be a trace gas (e.g., $< 1$ ppm) on such a planet.


\citet{Yu_2021} and \citet{Tsai_2021} also simulated a 1 bar H$_2$-dominated atmosphere on K2-18b, but instead with a hot $\sim 600$ K rocky surface (i.e., no habitable ocean). They find that $\sim 0.1\%$ to 1\% CH$_4$ can accumulate alongside $\sim 1\%$ of CO$_2$ and CO. We have done similar simulations with the \emph{Photochem} code and found that larger CH$_4$ concentrations are stable in this case because the hot $\sim 600$ K surface breaks down the kinetic barriers to CH$_4$ production. For example, when temperature is increased from 320 K to 600 K, the rate of the reaction $\mathrm{HCO} + \mathrm{H_2} \rightarrow \mathrm{H_2CO} + \mathrm{H}$ increases by about six orders of magnitude while Reaction \eqref{eq:ch4_prod2_4} increases by a factor of $\sim 30$. In contrast, methane production is far more kinetically inhibited on a Hycean planet with a habitable 320 K surface (e.g., model 1). Furthermore, the 1 bar scenarios in \citet{Yu_2021} and \citet{Tsai_2021} with rocky surfaces can be ruled out because such a planet would be denser than K2-18b's observed density. To explain the planet's mass and radius with only a silicate interior and a H$_2$-rich envelope, interior modeling suggests the atmosphere needs to be $\gtrsim 1000$ bars thick \citep{Madhusudhan_2020,Madhusudhan_2023}.

\subsection{Can CH$_4$ accumulate from non-photochemical abiotic processes?}

Substantial methane from non-photochemical abiotic processes is hard to sustain on a Hycean planet. To explain K2-18b's density, a Hycean world needs a large high-pressure ice layer that separates deep rocky material from the surface water ocean \citep{Madhusudhan_2021}. Water-rock reactions and subsequent transport of CH$_4$ is conceivable \citep{Thompson_2022}, but improbable in this case since the high overburden pressure of water and ice inhibits the production of fresh crust to be hydrated \citep{KrissansenTotton_2021}. Moreover, the shutdown of melting of deep-subsurface silicates also precludes the possibility of volcanic CH$_4$ \citep{KrissansenTotton_2021,Noack_2016,Kite_2018}. Massive asteroid impacts on the early Earth may have made transient atmospheric methane \citep{Wogan_2023}. However, ephemeral impact-induced CH$_4$ is unlikely on K2-18b because the planet is $\sim 2-3$ billion years old \citep{Guinan_2019}, while substantial bombardment is expected to end within the first several hundred million years of planet formation \citep{Lichtenberg_2022}.

\subsection{Inhabited Hycean vs. mini-Neptune: evaluating model complexity} \label{sec:discussion_hy_vs_nep}

Our results suggest that both an inhabited Hycean world (model 2) or a mini-Neptune with a massive H$_2$ atmosphere (model 3) are not strongly ruled out by the JWST data ($\chi_r^2 \approx 1.5$). However, in addition to evaluating the fit to the data, we also must assess the relative complexity of each scenario. The inhabited Hycean world (model 2) requires a cool habitable surface, but models suggest that a cloud-free $1$-bar H$_2$-rich atmosphere should trigger a hot runway greenhouse \citep[Figure \ref{fig:olr_isr},][]{Innes_2023,Pierrehumbert_2023}. A supercritical steam-dominated atmosphere would have a small scale height incompatible with JWST observations \citep{Scheucher_2020}. For a temperate surface, climate codes need to assume the presence of high-altitude clouds or hazes that scatter away starlight \citep{Madhusudhan_2021,Piette_2020}. 


Beyond this climate paradox, $\sim 1$ bar of H$_2$ may also be susceptible to rapid escape driven by extreme-ultraviolet radiation \citep[i.e., XUV,][]{Hu_2023}. Even if a 1 bar H$_2$ atmosphere could withstand modern XUV radiation, K2-18b likely experienced exceptionally high XUV fluxes during the host M star's pre-main sequence, potentially driving hundreds of bars of H$_2$ loss \citep{Luger_2015}, as so a remnant thin $\sim 1$ bar atmosphere would be highly fortuitous. As noted previously, replenishing H$_2$ with volcanism would be unlikely on a Hycean K2-18b because rocky material in the deep subsurface would be at pressures too high for melting and outgassing \citep{Kite_2018,Noack_2016}.

In contrast, our model of a gas-giant mini-Neptune (model 3) is relatively straightforward. For a $100 \times$ solar composition, solar C/O, and $T_\mathrm{int} = 60$ K, which are physically plausible given K2-18b's mass, a spectrum broadly consistent with the JWST data falls out of our model. Unlike a Hycean world, a mini-Neptune does not require a biosphere to explain the disequilibrium combination of atmospheric CH$_4$ and CO$_2$. Instead, these gases emerge in model 3 from deep-atmosphere quenching (Figure \ref{fig:neptune_mix_t_p}). Even though both an inhabited Hycean world and a mini-Neptune are allowed by JWST data, the climate of a Hycean world and the atmosphere's resilience to escape is hard to explain, so we favor the mini-Neptune model for its simplicity.

While our mini-Neptune simulation is relatively simple, it makes assumptions that should be investigated with more sophisticated modeling. Namely, we simulate the planet's climate (Figure \ref{fig:neptune_mix_t_p}) using a two-stage approach that is not fully self-consistent with photochemistry (Section \ref{sec:methods_neptune}), yet an accurate tropopause temperature is important for predicting whether H$_2$O cold-trapping can reproduce the JWST non-detection of water vapor. The water vapor cold trap would be better approximated by a model that is self-consistent with photochemistry and accounts for the possibility of convection inhibition \citep{Innes_2023}. The need for a H$_2$O cold trap is not unique to a mini-Neptune planet. A Hycean world would also need substantial water condensation to explain the H$_2$O non-detection. An additional shortcoming of this study is that we only consider one mini-Neptune scenario with a composition of $100\times$ solar metallicity, and solar C/O. Further modeling could tune metallicity and the  C/O ratio to get an even better fit to the JWST observations.

\subsection{Future observations of K2-18b}

Clearly distinguishing between the inhabited Hycean and mini-Neptune interpretations with future JWST observations will be challenging. Ammonia should be unique to a mini-Neptune K2-18b \citep[Figure \ref{fig:neptune_mix_t_p},][]{Yu_2021,Tsai_2021,Madhusudhan_2023_chem,Hu_2021}. However, even if future observations are unable to detect NH$_3$, this would not necessarily prove the inhabited Hycean case. In our mini-Neptune model, the NH$_3$ features act to fill CH$_4$ spectral windows (Figure \ref{fig:spectra_clear_1}). This small ammonia absorption is difficult to distinguish from clouds which have a similar effect on the spectrum (Figure \ref{fig:spectra_cloud}). Additionally, there are several reasons why ammonia could be less abundant on a mini-Neptune K2-18b than we have estimated, making a detection even more challenging. \citet{Hu_2021_gas} predicted stratospheric NH$_3$ could be photochemically depleted to undetectable concentrations ($< 1$ ppm) on a gas-rich K2-18b if tropospheric mixing is slow ($\lesssim 10^3$ cm$^2$ s$^{-1}$). Also, nitrogen could dissolve into a magma ocean at the base of a thick H$_2$-rich envelope, preventing a large \hl{observable} NH$_3$ abundance in the upper atmosphere \citep{Oliver_2024}.


An inhabited Hycean world could be identified with the detection of a biogenic gas. \citet{Madhusudhan_2023} found weak evidence for dimethyl sulfide (DMS) in K2-18b's transmission spectrum, a gas almost exclusively produced by life on Earth \citep{Catling_2018}. If DMS is detected with statistical significance, it might be difficult to account for its presence without a biosphere on a Hycean planet.

\section{Conclusions}

Recent JWST observations of K2-18b \citep{Madhusudhan_2023}, a habitable-zone sub-Neptune exoplanet, revealed the presence of atmospheric CH$_4$ and CO$_2$. \citet{Madhusudhan_2023} suggested that the data are best explained by a habitable ``Hycean'' world.  Our photochemical and climate simulations of K2-18b as a lifeless Hycean world suggest such a planet would have ppb-level CH$_4$ because the gas is rapidly destroyed by photolysis and subsequent oxidizing reactions. Lacking substantial CH$_4$, an uninhabited Hycean planet cannot explain these recent JWST observations, which suggest $\sim 1\%$ of the gas is present \citep{Madhusudhan_2023}. However, there are still two scenarios that fit the JWST observations equally well according to a $\chi_r^{2}$ metric: a Hycean world inhabited by methanogenic life, or a mini-Neptune with no defined surface. The latter case is less complex and requires fewer assumptions. 

Specifically, an inhabited Hycean K2-18b has the following difficulties:

\begin{itemize}
    \item To explain the $\sim 1\%$ CH$_4$ detected by JWST, a Hycean planet needs biogenic CH$_4$ or some other unknown source of the gas to maintain it against photochemical destruction. 
    \item Models predict that a stable temperate climate is challenging on a Hycean K2-18b. Such a planet is expected to experience a steam runaway greenhouse \citep[Figure \ref{fig:olr_isr};][]{Innes_2023,Scheucher_2020,Pierrehumbert_2023}, unless starlight can be reflected away by clouds \citep{Madhusudhan_2021,Piette_2020}.
    \item A thin $\sim 1$ bar H$_2$ atmosphere may be susceptible to XUV-driven escape. H$_2$ can not be replenished by volcanism because the overburden pressure of the thick ice and ocean layer on a Hycean world would prevent silicate melting \citep{Noack_2016,Kite_2018}. 
\end{itemize}
On the other hand, the benefits of the mini-Neptune case are: 

\begin{itemize} 
    \item The CH$_4$ and CO$_2$ detected by JWST can be broadly explained by deep-atmosphere thermochemical quenching for a $100\times$ solar metallicity, solar C/O, and 60 K intrinsic temperature.
    \item Deep atmosphere kinetics also predicts NH$_3$ and CO abundances generally compatible with the JWST non-detections of each gas.
    \item The lack of H$_2$O features in the spectrum can be accounted for by water vapor condensation and cold-trapping.
    \item Basic 1-D radiative-convective-equilibrium modeling can explain the planet's climate.
\end{itemize}

Overall, we favor the mini-Neptune explanation of K2-18b because it is simple and has fewer challenges than a Hycean interpretation.

\section*{Acknowledgements}

We thank our anonymous reviewer who improved the quality of this article. Also, we thank Stephen Klippenstein for sharing unpublished reaction rate calculations and improving out understanding of methane kinetics. This work benefited from discussions with Giada Arney, Eddie Schwieterman, Victoria Meadows, Jacob Lustig-Yaeger, Tyler Robinson and Michaela Leung. N.F.W was supported by the NASA Postdoctoral Program. N.E.B acknowledges support from NASA’S Interdisciplinary Consortia for Astrobiology Research (NNH19ZDA001N-ICAR) under award number 19-ICAR19\_2-0041. S.-M.T. acknowledges support from NASA Exobiology Grant No. 80NSSC20K1437. R.H. was supported in part by NASA Exoplanets Research Program grant \#80NM0018F0612. The research was carried out in part at the Jet Propulsion Laboratory, California Institute of Technology, under a contract with the National Aeronautics and Space Administration.

\section*{Software}

The source code needed to install the necessary software and reproduce all main text calculations (i.e., Figures \ref{fig:olr_isr} to \ref{fig:spectra_clear_1}) is archived on Zenodo: \url{https://doi.org/10.5281/zenodo.10537133} \citep{Wogan_2024_reproduce}.

\appendix

\renewcommand{\thefigure}{A\arabic{figure}}
\renewcommand{\theHfigure}{A\arabic{figure}}
\renewcommand{\thetable}{A\arabic{table}}
\renewcommand{\theHtable}{A\arabic{table}}

\setcounter{figure}{0}
\setcounter{table}{0}

\section{Reaction rate updates and the mini-Neptune P-T-$K_{zz}$ profile}

\begin{figure*}
  \centering
  \includegraphics[width=0.75\textwidth]{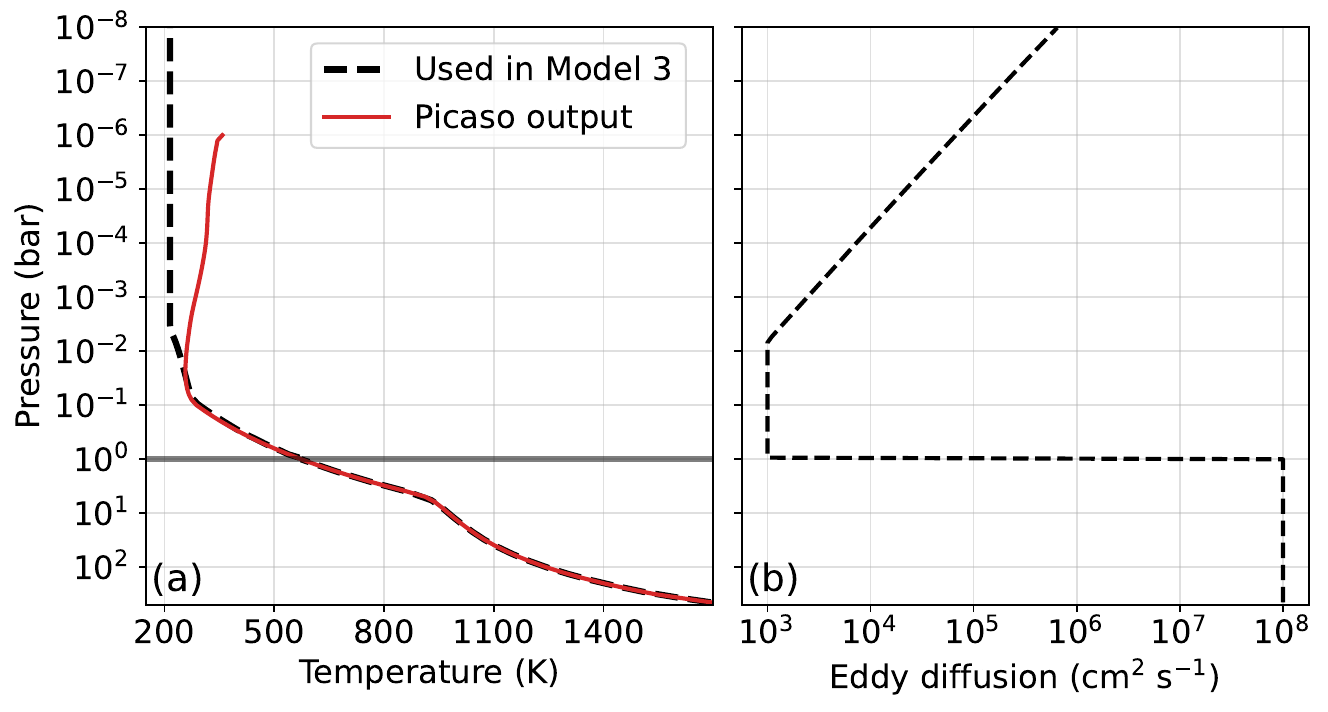}
  \caption{(a) shows the P-T profile for a mini-Neptune K2-18b (model 3). We compute the red P-T profile using the \emph{PICASO} climate model. As described in Section \ref{sec:methods_neptune}, we modify the \emph{PICASO} result to make the black dashed P-T profile, which we use in model 3 (Table \ref{tab:models}). (b) is the assumed eddy diffusion coefficient for model 3 adopted from \citet{Hu_2021_gas}.}
  \label{fig:neptune_clima_comp}
\end{figure*}

Table \ref{tab:updated_reactions} archives the chemical reactions in our network that we updated for this study. Our updated branching ratios for H$_2$O photolysis are listed in Table \ref{tab:h2o_branch} based on \citet{Slanger_1982} and \citet{Stief_1975}. Figure \ref{fig:neptune_clima_comp} illustrates our computed P-T profile for a mini-Neptune K2-18b using the \emph{PICASO} code compared to the modified P-T profile we use in model 3 (see Section \ref{sec:methods_neptune} for details). The figure also shows our assumed $K_{zz}$ profile, which we adopted from \citet{Hu_2021_gas}.

\section{Clouds and hazes}

Here, we consider the effects of clouds and hazes on our K2-18b simulations. In both model 1 and 2, water vapor condenses from the surface to about 0.03 bar forming a cloud deck. Model 3 may also have water vapor clouds caused by condensation between 0.07 bar to 4 mbar. Our photochemical model predicts that hydrocarbon aerosols, similar to Titan's, are produce in model 2 and 3 at high altitudes (e.g., $10^{-5}$ bar) because both atmospheres have abundant CH$_4$. Finally, in model 3, photochemistry processes H$_2$S to elemental sulfur which condenses to a haze in the same region as the water cloud \citep{Zahnle_2016}.

\begin{figure*}
  \centering
  \includegraphics[width=1\textwidth]{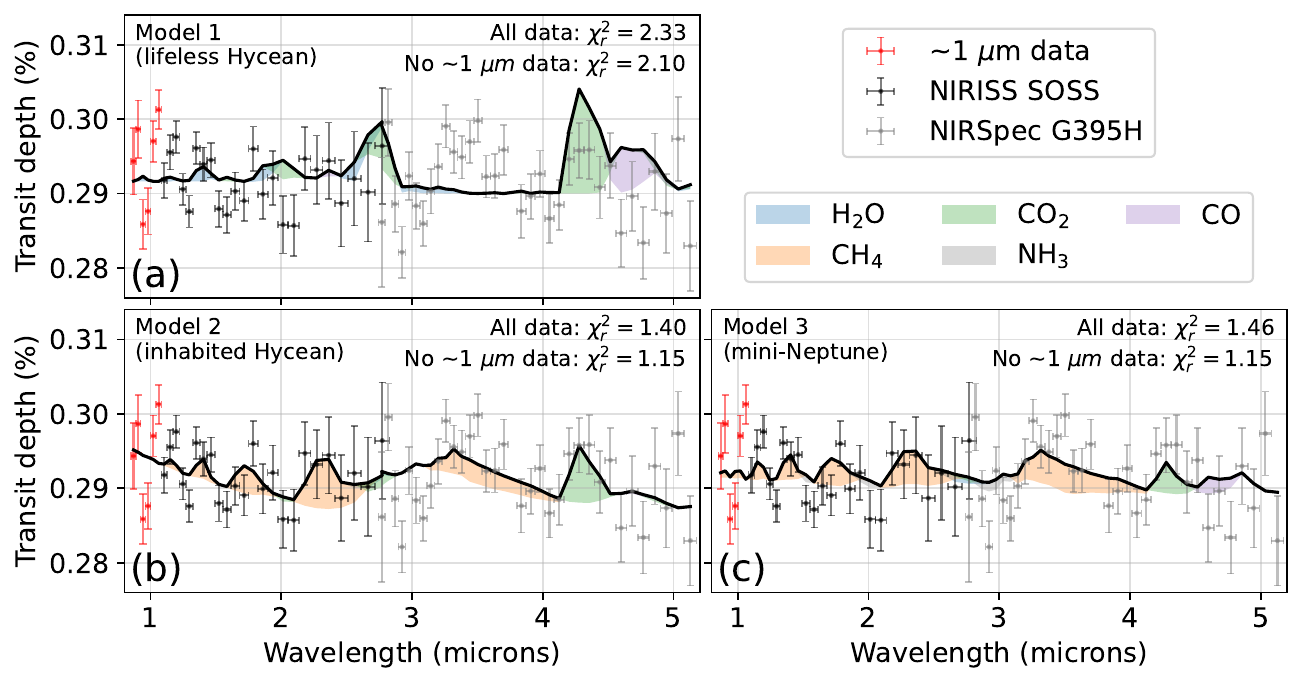}
  \caption{Similar to Figure \ref{fig:spectra_clear_1}, except all spectral calculations include clouds made of condensed water, elemental sulfur, and hydrocarbons. Each panel reports two $\chi^2_r$ values: one accounts for all the JWST data, while the other excludes the six red points near $1$ $\mu$m to show the effect of this scatter on the $\chi^2_r$.}
  \label{fig:spectra_cloud}
\end{figure*}

Figure \ref{fig:spectra_cloud} shows simulated spectra of models 1, 2 and 3 that account for these clouds. The calculation uses a range of opacities appropriate for each aerosol. For hydrocarbon aerosols, we adopt real and imaginary indexes of refraction appropriate for a Titan-like haze \citep{Khare_1984}. Condensed elemental sulfur has the optical properties shown in Figure S1 of \citet{Tian_2010}. These indexes of refraction only extent from 0.15 to 0.8 $\mu$m, so, following \citet{Hu_2021_gas}, we constantly extrapolate to longer wavelengths. For both sulfur clouds and hydrocarbon hazes, we use particle densities predicted by the \emph{Photochem} model and assume all aerosols are perfect Mie spheres with a 0.1 $\mu$m radius, the size being motivated by the particle radii in Titan's haze \citep{Rages_1983}. Calculations approximate water clouds by simply adding an opaque cloud layer wherever H$_2$O condenses in the atmosphere.

Overall, our conclusion in the main text that a lifeless Hycean planet (model 1) is ruled out by the JWST data remains unchanged when considering cloudy spectra (Figure \ref{fig:spectra_cloud}a). Figure \ref{fig:spectra_cloud}b and \ref{fig:spectra_cloud}c shows that, like in our clear-sky simulations (Figure \ref{fig:spectra_clear_1}), the data do not exclude the inhabited Hycean (model 2) or mini-Neptune (model 3) scenarios. Furthermore, the figure reports two $\chi_r^2$ values for each panel: one that includes all the JWST data, and another that excludes the six red data points near 1 $\mu$m. This shows that the data scatter near 1 $\mu$m has a large effect on the $\chi_r^2$.

\begin{table*}
  \caption{Updated reaction rates and thermodynamics}
  \label{tab:updated_reactions}
  \begin{center}
  \begin{tabularx}{.9\linewidth}{p{0.27\linewidth} p{0.34\linewidth} p{0.25\linewidth} }
    \hline \hline
    Reaction & Rate$^\text{a}$ & Reference \\
    \hline
    $\mathrm{H} + \mathrm{H_2CO} + \mathrm{M} \rightarrow \mathrm{CH_3O} + \mathrm{M}$$^\text{b}$ & 
    $k_0 = 1.22 \times 10^{-23} T^{-3} \exp(-2900/T)$ \newline 
    $k_{\infty} = 6.56 \times 10^{3} T^{-5} \exp(-4000/T)$ &
    \citet{Xu_2015}
    \\
    $\mathrm{H} + \mathrm{H_2CO} + \mathrm{M} \rightarrow \mathrm{CH_3O} + \mathrm{M}$$^\text{b}$ & 
    $k_0 = 9.26 \times 10^{-23} T^{-3.38} \exp(-1432.7/T)$ \newline 
    $k_{\infty} = 2.65 \times 10^{-2} T^{-3.36} \exp(-2771.4/T)$ &
    \citet{Klippenstein_priv_com}
    \\
    $\mathrm{H} + \mathrm{H_2CO} + \mathrm{M} \rightarrow \mathrm{H_2COH} + \mathrm{M}$$^\text{b}$ & 
    $k_0 =  2.82 \times 10^{-29} T^{-1.2} \exp(-2900/T)$ \newline 
    $k_{\infty} = 3 \times 10^{-12} \exp(-3500/T)$ &
    \citet{Xu_2015}
    \\
    $\mathrm{H} + \mathrm{H_2CO} + \mathrm{M} \rightarrow \mathrm{H_2COH} + \mathrm{M}$$^\text{b}$ & 
    $k_0 =  2.99 \times 10^{-21} T^{-3.4} \exp(-2127.5/T)$ \newline 
    $k_{\infty} = 1.92 \times 10^{-25} T^{3.89} \exp(-516.9/T)$ &
    \citet{Klippenstein_priv_com}
    \\
    $\mathrm{CH_3} + \mathrm{O} \rightarrow \mathrm{H_2CO} + \mathrm{H}$ &
    $9 \times 10^{-11}$ &
    \citet{Xu_2015}
    \\
    $\mathrm{CH_3} + \mathrm{O} \rightarrow \mathrm{HCO} + \mathrm{H_2}$ &
    $6 \times 10^{-11}$ &
    \citet{Xu_2015}
    \\
    $\mathrm{H_2CO} + \mathrm{H} \rightarrow \mathrm{HCO} + \mathrm{H_2}$ &
    $2.28 \times 10^{-19} T^{2.65} \exp(-766.5/T)$ &
    \citet{Xu_2015}
    \\
    $\mathrm{H_2O} + h\nu \rightarrow \mathrm{OH} + \mathrm{H}$ &
    Determined by photolysis cross section$^\text{c}$ &
    \citet{Slanger_1982,Stief_1975}
    \\
    $\mathrm{H_2O} + h\nu \rightarrow \mathrm{H_2} + \mathrm{O(^1D)}$ &
    Determined by photolysis cross section$^\text{c}$ &
    \citet{Slanger_1982,Stief_1975}
    \\
    $\mathrm{H_2O} + h\nu \rightarrow \mathrm{O} + \mathrm{H} + \mathrm{H}$ &
    Determined by photolysis cross section$^\text{c}$ &
    \citet{Slanger_1982,Stief_1975}
    \\
    $\mathrm{H_2} + h\nu \rightarrow \mathrm{H} + \mathrm{H}$ &
    Determined by photolysis cross section &
    \citet{Heays_2017}
    \\
    \hline \hline
    Species & Enthalpy$^\text{d}$ (KJ/mol) & Reference \\
    \hline
    $\mathrm{CH_3O}$ & 21.6 &
    \citet{Xu_2015} \\
    $\mathrm{H_2COH}$ & -15.3 &
    \citet{Xu_2015} \\
    \hline
    \multicolumn{3}{p{0.9\linewidth}}{
      $^\text{a}$Low pressure rate constants, $k_0$, have units cm$^6$ molecules$^{-2}$ s$^{-1}$. All other rates have units cm$^3$ molecules$^{-1}$ s$^{-1}$.

      $^\text{b}$Reactions $\mathrm{H} + \mathrm{H_2CO} + \mathrm{M} \rightarrow \mathrm{CH_3O} + \mathrm{M}$ and $\mathrm{H} + \mathrm{H_2CO} + \mathrm{M} \rightarrow \mathrm{H_2COH} + \mathrm{M}$ have two rate entries. We nominally adopt the rate from \citet{Xu_2015} but also consider the \citet{Klippenstein_priv_com} rate as a sensitivity test (Section \ref{sec:result_hycean}).

      $^\text{c}$We updated the branching ratios for these three reactions, but not the total photolysis cross section.

      $^\text{d}$Enthalpy of formation at 298 K.
    }
  \end{tabularx}
  \end{center}
\end{table*}

\begin{table*}
  \caption{Updated H$_2$O photolysis branching ratios}
  \label{tab:h2o_branch}
  \begin{center}
  \begin{tabularx}{.62\linewidth}{p{0.15\linewidth} p{0.12\linewidth} p{0.12\linewidth} p{0.12\linewidth}}
    \hline \hline
    Wavelength$^\text{a}$ (nm) & $\mathrm{OH} + \mathrm{H}$ & $\mathrm{H_2} + \mathrm{O(^1D)}$ & $\mathrm{O} + \mathrm{H} + \mathrm{H}$ \\
    \hline
    92.5 & 0.89 & 0.11 & 0 \\
    120.9 & 0.89 & 0.11 & 0 \\
    121.0 & 0.78 & 0.1 & 0.12 \\
    122.1 & 0.89 & 0.11 & 0 \\
    145.0 & 1 & 0 & 0 \\
    251.6 & 1 & 0 & 0 \\
    \hline
    \multicolumn{4}{p{0.62\linewidth}}{
      $^\text{a}$Branching ratios are linearly interpolate to intermediate wavelengths.
    }
  \end{tabularx}
  \end{center}
\end{table*}

\bibliography{bib}
\bibliographystyle{aasjournal}

\end{document}